# Annual Benefit Analysis of Integrating the Seasonal Hydrogen Storage into the Renewable Power Grids

Jin Lu
*Student Member, IEEE*
Department of Electrical and Computer Engineering
University of Houston
Houston, TX, USA
jlu28@uh.edu

Xingpeng Li
*Senior Member, IEEE*
Department of Electrical and Computer Engineering
University of Houston
Houston, TX, USA
xli82@uh.edu

*Abstract*—There has been growing interest in integrating hydrogen storage into power grids with high renewable penetration levels. The economic benefits and power grid reliability are both essential for hydrogen storage integration. In this paper, an annual scheduling model (ASM) for energy hubs (EH) coupled power grids is proposed to investigate the annual benefits of seasonal hydrogen storage (SHS). Each energy hub consists of hydrogen storage, electrolyzers, and fuel cells. The electrical and hydrogen energy can be exchanged on the bus with the energy hub. The physical constraints for both grids and EHs are enforced in ASM. The proposed ASM considers the intra-season daily operation of the EH coupled grids. Four typical daily profiles are used in ASM to represent the grid conditions in four seasons, which reduces the computational burden. Besides, both the intra-season and cross-season hydrogen exchange and storage are modeled in the ASM. Hence, the utilization of hydrogen storage is optimized on a year-round level. Numerical simulations are conducted on the IEEE 24-bus system. The simulation results indicate that seasonal hydrogen storage can effectively save the annual operation cost and reduce renewable curtailments.

## Nomenclature

*Indices*
| | |
|---|---|
| $e$ | Electrolyzer. |
| $f$ | Fuel cell. |
| $g$ | Generator. |
| $k$ | Branch. |
| $n$ | Bus. |
| $q$ | Quarter. |
| $t$ | Time period. |
| $w$ | Wind power plant. |

*Sets*
| | |
|---|---|
| $D$ | Set of days in a quarter. |
| $E(n)$ | Set of electrolyzers on bus $n$. |
| $F(n)$ | Set of fuel cells on bus $n$. |
| $G$ | Set of generators in the power system. |
| $G(n)$ | Set of generators on bus $n$. |
| $K$ | Set of branches in the power system. |
| $K(n+)$ | Set of branches that the starting bus is $n$. |
| $K(n-)$ | Set of branches that the ending bus is $n$. |
| $Q$ | Set of quarters in a year. |
| $T$ | Set of time periods in a day. |
| $T^D$ | Set of time periods in a day for previous daily stored hydrogen calculation. |
| $T^P(t)$ | Set of time periods in a day before period $t$. |
| $W$ | Set of wind power plants in the power system. |
| $W(n)$ | Set of wind power plants located at bus $n$. |

*Parameters*
| | |
|---|---|
| $c_g$ | Operational cost for generator $g$. |
| $c_g^{NL}$ | No load cost for generator $g$. |
| $c_g^{SU}$ | Startup cost for generator $g$. |
| $d_{nqt}$ | Demand on bus $n$ at period $t$ in quarter $q$. |
| $P_e^{max}$ | Maximum capacity of electrolyzer $e$. |
| $P_f^{max}$ | Maximum capacity of fuel cell $f$. |
| $P_g^{max}$ | Maximum capacity of generator $g$. |
| $P_g^{min}$ | Minimum output power of generator $g$. |
| $P_k^{max}$ | Maximum thermal capacity of branch $k$. |
| $P_{wqt}$ | Available wind active power at period $t$ for quarter $q$. |
| $R_g^{10}$ | Outage ramping limit in 10 minutes for generator $g$. |
| $R_g$ | Ramping limit in an hour for generator $g$. |
| $x_k$ | Reactance of branch $k$. |
| $\eta_e$ | Energy conversion efficiency of electrolyzer $e$. |
| $\eta_f$ | Energy conversion efficiency of fuel cell $f$. |

*Variables*
| | |
|---|---|
| $E_{n,q}^0$ | The initial hydrogen energy stored at bus $n$ in quarter $q$. |
| $E_{nqtd}$ | The hydrogen energy stored at bus $n$ at period $t$ in day $d$ for quarter $q$. |
| $P_{eqt}$ | Active power consumed by electrolyzer $e$ at period $t$ for quarter $q$. |
| $P_{fqt}$ | Active power generated by fuel cell $f$ at period $t$ for quarter $q$. |
| $P_{gqt}$ | The output power of generator $g$ at period $t$ for quarter $q$. |
| $P_{kqt}$ | Active power flow on branch $k$ at period $t$ for quarter $q$. |
| $P_{wqt}^{Cur}$ | Active power curtailment for wind power plant $w$ at period $t$. |
| $r_{gqt}$ | Reserve from generator $g$ in period $t$ for quarter $q$. |
| $u_{gqt}$ | Commitment status for generator $g$ in period $t$ for quarter $q$. |
| $v_{gqt}$ | Generator startup indicator, 1 if generator $g$ is turned on in period $t$; 0 otherwise. |
| $\theta_{kqt}$ | Phase angle of the branch $k$ at period $t$ for quarter $q$. |

## I. Introduction

The modern power grids are incorporating more renewable resources. Although renewable energy can help the grids become cleaner and greener, some issues arise due to the high penetration level of renewables [1]-[3]. Specifically, the renewables such as wind and solar are intermittent and more flexible resources with fast ramping rate are required when renewable generation drop rapidly in a short time period. Besides, renewable resources may not locate near the loads. The renewable curtailment is frequently observed in the renewable power grids with limited transmission capability. Many studies show that energy storage can enhance the performance of the grids, especially when the grids have a high renewable penetration level. Batteries, electric vehicles, and other types of energy storage are studied to incorporate with the grid and mitigate the issues caused by renewables. Among various energy storage, hydrogen storage is regarded as one of the promising technologies. Hydrogen can be stored by using different technologies such as gas cylinders, liquid tanks, absorptive materials, interstitial sites in host metal, etc.[4]. The large capacity hydrogen storage is required for grid-scale operation [5]. In [6], the methods and design aspects of salt caverns are investigated to realize large-scale hydrogen storage. In [7]-[9], researchers investigate the potential of practically applying salt caverns in energy systems.

To integrate hydrogen storage in the power grids, the power to hydrogen (P2H) and hydrogen to power (H2P) facilities are required for the electrical and hydrogen energy exchange [10]-[12]. P2H facilities are used to generate hydrogen by consuming electrical energy, and the produced hydrogen can be stored in hydrogen storage. H2P facilities can generate electrical energy by consuming hydrogen, which means the stored hydrogen energy can be converted back to electrical energy.

Several studies propose different hydrogen storage integration methods in power grids. In [13], it is proposed to couple the power grids with energy hubs (EHs). Each EH contains the P2H, H2P, and hydrogen storage at the same location. In [14], the hydrogen can be transmitted through the pure hydrogen / hydrogen mixed pipelines. Both [13] and [14] focus on daily operation. Very few studies in literature investigate long-term management/operation. In [15], the mathematical model for seasonal hydrogen storage in the multi-energy system is studied. In [16], the annual management of power grids coupled with the hydrogen and heat systems are studied.

In this paper, we proposed an annual scheduling model (ASM) which can schedule the optimal daily hydrogen exchange operations for different seasons of the whole year. To achieve this goal, the intra-day electrical-hydrogen energy exchange constraints are included in the typical daily operation of the ASM. Hence, the fluctuation of the renewables such as solar power can be mitigated. Besides, the cross-season electrical-hydrogen exchange can be captured using the accumulated hydrogen storage formula in the ASM. Hence, the optimized EH operations can better utilize the renewable resources while the loads and renewables vary between different seasons.

The rest of this paper contains four sections. The EH coupled power grids and the proposed ASM are presented in section II. Section III discusses the benchmark model, which is the annual scheduling model for the power grids without EHs. Section IV includes case studies and simulation result analysis. Section V is the conclusion.

## II. ASM OF EH COUPLED GRIDS

In this section, the power grids coupled with EH are introduced and the modeling of its annual operation is discussed.

### A. EH Coupled Renewable Power Grids

In a power grid coupled with EHs, the electrical and hydrogen energy can be exchanged at the buses where EHs locate. In the aspect of power grid operation, the extra electrical energy can be stored as hydrogen energy and converted back to electrical energy when needed. In the power grids with high renewable penetration levels, there will be extra electrical energy at the buses with high renewable generation. The EHs located at these buses can store the extra energy in hydrogen storage. Hence, it is expected to mitigate the issues caused by the intermittence of the renewables.

The loads are influenced by the temperature and other environmental variables. For example, the loads in summer are much higher than the loads in winter in ERCOT [17]. Since hydrogen storage can store energy for long periods, the EHs can store hydrogen energy in low-load seasons and release electrical energy in high-load seasons.

Based on the above reasons, the EH coupled renewable power grids should include the electrical-hydrogen exchange constraints for both short and long time periods. In this paper, the annual scheduling model of the EH coupled grid is proposed and described in the following subsection.

### B. Modeling of Annual operation for EH Coupled Grid

To consider the short time period electrical-hydrogen exchange operations, the proposed ASM uses four typical days to describe the daily operations in four quarters. Specifically, we assume the renewable generation and load profiles are the same for all days in a quarter. Hence, the daily power system operations remain the same for all days in the quarter. The daily operation constraints considering different quarters are listed as follows. The generator power maximum and minimum power limits are shown in Eq. (1)-(2). Besides, the generator reserve's ramping limit and minimum reserve constraint are shown in (3)-(4). Eq. (5) is the DC power flow equation, and Eq. (6) represents the transmission line power limits. Since this study focuses on grids with high renewable penetration level, the curtailment of renewables is modeled using Eq. (7). The generator ramping limit is represented by Eq. (8). The relationship of the two binary variables $v_{gt}$ and $u_{gt}$ is shown in Eq. (9).

$$P_g^{min} u_{gqt} \leq P_{gqt} \quad \forall g, q, t \tag{1}$$

$$P_{gqt} + r_{gqt} \leq P_g^{max} u_{gqt} \quad \forall g, q, t \tag{2}$$

$$0 \leq r_{gqt} \leq R_g^{10} u_{gqt} \quad \forall g, q, t \tag{3}$$

$$\sum_{m \in G} r_{mqt} \geq P_{gqt} + r_{gqt} \quad \forall g, q, t \tag{4}$$

$$P_{kqt} = \theta_{kqt}/x_k \quad \forall k, q, t \tag{5}$$

$$-P_k^{max} \leq P_{kqt} \leq P_k^{max} \quad \forall k, q, t \tag{6}$$

$$0 \leq P_{wqt}^{Cur} \leq P_{wt} \quad \forall w, q, t \tag{7}$$

$$-R_g \leq P_{gqt} - P_{g,q,t-1} \leq R_g \quad \forall g, q, t \geq 2 \tag{8}$$

$$v_{gqt} \geq u_{gqt} - u_{g,q,t-1} \quad \forall g, q, t \geq 2 \tag{9}$$

$$v_{gqt} \in \{0,1\} \quad \forall g, q, t \tag{10}$$

$$u_{gqt} \in \{0,1\} \quad \forall g, q, t \tag{11}$$

The nodal power balance equation in the EH coupled grids should include the electrical power generated/consumed by the fuel cells/electrolyzers. Besides, renewable generation and curtailment are also included in the equation. The nodal power balance equation is shown in Eq. (12).

$$\sum_{g\in G(n)} P_{gqt} + \sum_{k\in K(n-)} P_{kqt} - \sum_{k\in K(n+)} P_{kqt} + \sum_{w\in W(n)} P_{wqt}$$
$$- \sum_{w\in W(n)} P_{wqt}^{Cur} + \sum_{f\in F(n)} P_{fqt} = d_{nqt} + \sum_{e\in E(n)} P_{eqt} \quad \forall n,q,t \quad (12)$$

The fuel cells and electrolyzers have maximum power capacity, which is described by Eq. (13)-(14).

$$0 \leq P_{eqt} \leq P_e^{max} \quad \forall e,t \quad (13)$$
$$0 \leq P_{fqt} \leq P_f^{max} \quad \forall f,t \quad (14)$$

The same daily operation will repeat for all days in the quarter. Following constraints involved the last hour of the previous day and the first hour of the current day. Eq. (15) is the generator ramping rate constraint between the days in the same quarter. Eq. (16) is the binary variable constraint for the two connecting hours between the days.

$$-R_g \leq P_{g,q,1} - P_{g,q,24} \leq R_g \quad \forall g,q \quad (15)$$
$$v_{g,q,1} \geq u_{g,q,1} - u_{g,q,24} \quad \forall g,q \quad (16)$$

To consider the seasonal electrical-hydrogen exchange operation, the stored hydrogen at different quarters is modeled. The stored hydrogen at the specific hour and day can be calculated per Eq. (17). $E_{nqtd}$ is the hydrogen energy stored at bus $n$ at period $t$ for day $d$ in Quarter $q$. $E_{n,q}^0$ is the initial hydrogen energy stored at bus $n$ for Quarter $q$. The second addition term on the right-hand side is the total hydrogen energy generated/consumed in the previous days of the quarter. $T^D$ is the set of all hours in a day. $P_{eqt'}$ and $P_{fqt'}$ are the power generated/consumed by the electrolyzer and fuel cell. The third addition term on the right-hand side calculates the accumulated hydrogen energy generated/consumed in the previous hours of the day. $T^P$ is the set of hours in a day which are earlier than hour $t$.

$$E_{nqtd} = E_{n,q}^0 + \sum_{t'\in T^D}\left(\sum_{e\in E(n)} \eta_e P_{eqt'} - \sum_{f\in F(n)} P_{fqt'}/\eta_f\right) * (d-1)$$
$$+ \sum_{t''\in T^P(t)}\left(\sum_{e\in E(n)} \eta_e P_{eqt''} - \sum_{f\in F(n)} P_{fqt''}/\eta_f\right)$$
$$\forall g,q,t,d \quad (17)$$

The initial stored hydrogen energy in a quarter equals the stored hydrogen energy in the last hour of the last day in the previous quarter. It is described by Eq. (18). Besides, we assume the stored hydrogen energy remains the same after the one-year operation, which is shown in Eq. (19).

$$E_{n,q}^0 = E_{n,q-1,24,90} \quad \forall n, q \geq 2 \quad (18)$$
$$E_{n,4,24,90} = E_{n,1}^0 \quad \forall n \quad (19)$$

With all the above constraints, the objective of the annual operation is to minimize the total operation cost including generator operational cost, startup cost, and no-load cost. We assume each quarter contains 90 days. The objective function is represented using Eq. (20).

$$\min \sum_{g\in G}\sum_{q\in Q}\sum_{t\in T}(c_g P_{gqt} + c_g^{NL} u_{gqt} + c_g^{SU} v_{gqt}) * 90 \quad (20)$$

Based on the above statement, the ASM for EH coupled grids includes Eq. (1)-(20).

### III. BENCHMARK MODEL FOR GRIDS WITHOUT EHs

To analyze the performance and benefits of the EH coupled power grids using ASM, we also develop the ASM for the traditional grids without EHs. Compared to EH coupled grids, traditional grids do not have electrical-hydrogen exchange. Hence, the difference between the ASM for EH coupled grids (EH-ASM) and the ASM for traditional grids (T-ASM) is that no hydrogen related constraints are included in the T-ASM. The nodal power balance equation without P2H and H2P facilities is shown in Eq. (21).

$$\sum_{g\in G(n)} P_{gqt} + \sum_{k\in K(n-)} P_{kqt} - \sum_{k\in K(n+)} P_{kqt} + \sum_{w\in W(n)} P_{wqt}$$
$$- \sum_{w\in W(n)} P_{wqt}^{Cur} = d_{nqt} \quad \forall n,q,t \quad (21)$$

The T-ASM includes Eq. (1)-(11), (15)-(16), (20)-(21).

### IV. CASE STUDIES

The modified IEEE 24-bus system is selected as the test case. The quarterly loads and wind production profiles are created which can simulate the actual hourly fluctuation in four typical days. Besides, the four quarters' profiles also show the seasonal fluctuation. For example, the loads in quarter 3 are higher than in other quarters. For EH coupled case, two energy hubs are located on bus 14 and bus 22. Since the wind farms sit on these two buses, the wind energy can convert into the hydrogen energy and be stored. Each energy hub contains a pair of fuel cell and electrolyzer. The efficiency of fuel cells $\eta_f$ is set to 0.6, and the efficiency of electrolyzers $\eta_e$ is set to 0.8. The hydrogen storage at each energy hub has 5000 MWh maximum energy capacity. The benchmark case for T-ASM simulation is the same 24-bus system without EHs.

The proposed EH-ASM and benchmark T-ASM are implemented using Python with the pyomo package and GLPK solver. The simulation is performed on a desktop computer with Intel-i7 3.2 GHz CPU and 16 GB RAM.

Two scenarios with 20% and 50% wind penetration levels are considered in this simulation. The quarterly wind production profiles for four typical days at different wind penetration levels are created. Then, the EH-ASM and T-ASM simulation is run for these wind penetration scenarios. The simulation results for EH-ASM and T-ASM at 20% wind penetration level are shown in Table I and Table II separately.

Table I. EH-ASM Simulation Result at 20% Wind Penetration Level

|  | Quarter 1 | Quarter 2 | Quarter 3 | Quarter 4 |
|---|---|---|---|---|
| Wind Curtailment (MWh) | 0 | 0 | 0 | 0 |
| Conventional Generation (MWh) | $3.52*10^6$ | $3.82*10^6$ | $4.8*10^6$ | $3.56*10^6$ |
| Average Power Flow Percentage (%) | 39.3% | 42.9% | 42.3% | 42.2% |
| Total Cost ($) | 258.53M | | | |

Table II. T-ASM Simulation Result at 20% Wind Penetration Level

|  | Quarter 1 | Quarter 2 | Quarter 3 | Quarter 4 |
|---|---|---|---|---|
| Wind Curtailment (MWh) | 0 | 0 | 0 | 0 |
| Conventional Generation (MWh) | $3.45*10^6$ | $3.67*10^6$ | $4.75*10^6$ | $3.45*10^6$ |
| Average Power Flow Percentage (%) | 41.0% | 40.2% | 42.4% | 41.2% |
| Total Cost ($) | 267.01M | | | |

As shown in tables I and II, neither the EH-ASM case nor the T-ASM case has to curtail any wind power at 20% wind penetration level. The conventional generation of the EH-ASM case is higher than the T-ASM case due to the losses in energy conversion between hydrogen and electricity. However, the total operation cost of the EH-ASM is about 3.17% lower than the T-ASM case. Since the proposed ASMs model the daily operation of the typical days, the hourly grid operation condition is also obtained from the ASM simulation. The hourly conventional generation at 20% wind penetration level for 4 typical days is shown in Figs. 1-4. The conventional generation curve of EH-ASM has fewer high value periods, which means the grid does not need to start more expensive generators to meet these high generation periods. The reason is that the EH-ASM can optimize the hourly electrical and hydrogen energy exchange operation to reduce the operation cost. It indicates that hydrogen storage can benefit the renewable power grid even when the renewable penetration level is not very high.

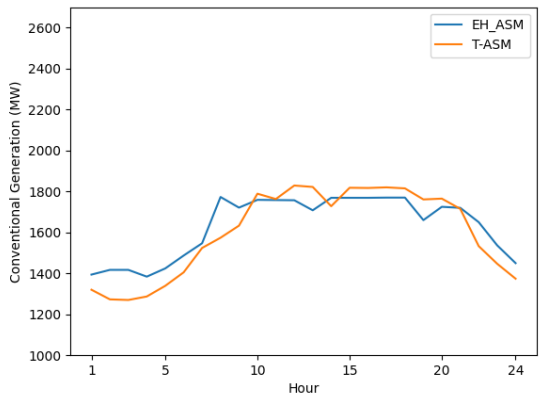

Fig. 1. The Conventional Generation in Quarter 1 at 20% Wind Penetration Level.

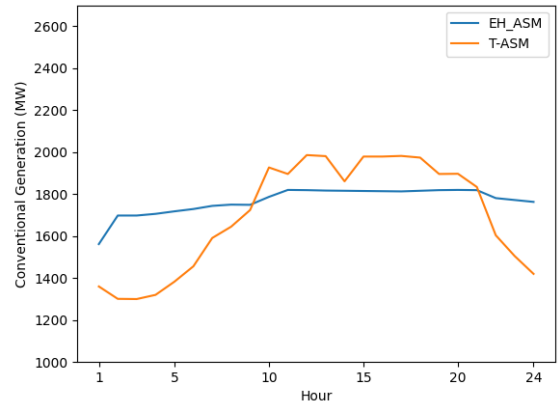

Fig. 2. The Conventional Generation in Quarter 2 at 20% Wind Penetration Level.

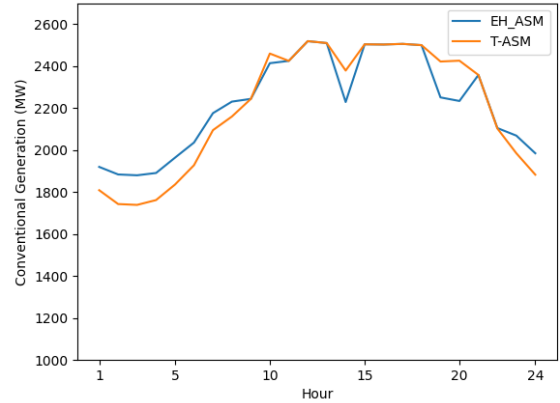

Fig. 3. The Conventional Generation in Quarter 3 at 20% Wind Penetration Level.

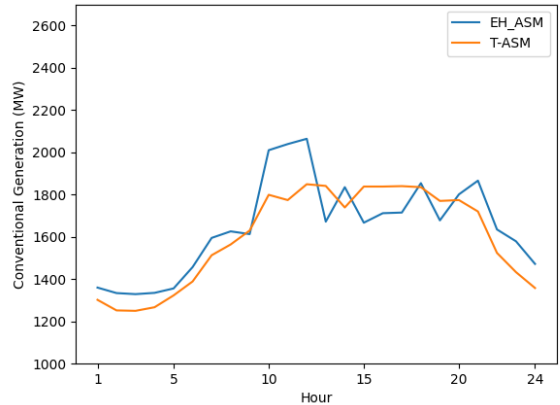

Fig. 4. The Conventional Generation in Quarter 4 at 20% Wind Penetration Level.

The simulation results for EH-ASM and T-ASM at 50% wind penetration level are shown in Tables III and IV. From these two tables, we can observe that the EH-ASM case has lower wind curtailment and total operation cost than the T-ASM case. Since the wind generation is large, the hydrogen storage cannot store all the wind energy. Therefore, wind curtailment still occurred in the EH-ASM case. However, the hydrogen storage can store the wind energy in the daily operation and reduce the conventional generation. Due to the above reasons, the performance of the EH-ASM case is better than the T-ASM case. To be noticed, the total hydrogen storage in the EH-ASM case is 10,000 MWh (two 5,000 MWh hydrogen storage on the EHs), which is the common salt cavern storage capacity. A larger hydrogen storage capacity may enhance the performance of the EH-ASM case.

Table III. EH-ASM Simulation Result at 50% Wind Penetration Level

|  | Quarter 1 | Quarter 2 | Quarter 3 | Quarter 4 |
|---|---|---|---|---|
| Wind Curtailment (MWh) | $2.13*10^5$ | $5.27*10^5$ | $7.15*10^5$ | $2.88*10^5$ |
| Conventional Generation (MWh) | $2.68*10^6$ | $3.01*10^6$ | $4.14*10^6$ | $2.52*10^6$ |
| Average Power Flow Percentage (%) | 38.2% | 41.5% | 41.3% | 39.6% |
| Total Cost ($) | 193.59M | | | |

Table IV. T-ASM Simulation Result at 50% Wind Penetration Level

|  | Quarter 1 | Quarter 2 | Quarter 3 | Quarter 4 |
|---|---|---|---|---|
| Wind Curtailment (MWh) | $2.17*10^5$ | $5.49*10^5$ | $7.33*10^5$ | $3.01*10^5$ |
| Conventional Generation (MWh) | $2.80*10^6$ | $3.09*10^6$ | $4.27*10^6$ | $2.80*10^6$ |
| Average Power Flow Percentage (%) | 37.6% | 40.0% | 40.8% | 39.9% |
| Total Cost ($) | 211.61M | | | |

In the above simulation, the efficiencies of electrolyzers and fuel cells are on the high end of their achievable range. The electrolyzer efficiency is set to 80% and fuel cell efficiency is set to 60%. The round-trip efficiency of electrical-hydrogen exchange is 48%. In more common cases, the round-trip efficiency of electrical-hydrogen exchange is about 37% [18]-[20]. Hence, the simulation for 37% round-trip efficiency at 50% wind penetration level is conducted. The total operational cost of EH-ASM is $195.37 million. For 48% round-trip efficiency, it is $193.59 million. Decreasing the round-trip efficiency from 48% to 37% increases the total cost by roughly 1%. It indicates that the round-trip efficiency has less impact on the total cost when wind penetration level is high.

## V. Conclusion

The proposed ASM considering electrical and hydrogen energy exchange operation for EHs integrated renewable power grids are developed in this work. The ASM uses four typical days to represent the system conditions of each quarter. For each typical day, the proposed ASM considers both the power system operation and the EH operation. Specifically, the electrical constraints for generation, transmission, and power balance are included; and the electrical-hydrogen exchange constraints related to fuel cell and electrolyzer are also enforced in the proposed ASM model. To realize the seasonal hydrogen storage/exchange, the stored hydrogen in each energy hub for each hourly time interval on each day in the year is calculated using the accumulative stored hydrogen equation. The proposed EH-ASM is tested on IEEE 24-bus system. The simulation results show that the proposed EH-ASM can mitigate both the daily and seasonal variations of the loads and renewable generation. Hence, the proposed EH-ASM can effectively reduce the annual operation cost and renewable curtailment during the year. This paper indicates that hydrogen storage can enhance the renewable power grid in the aspect of both daily and yearly operations.